# A Novel Vectorcardiogram System

Gabriel E. Arrobo[†], Calvin A. Perumalla[†], Yang Liu[†], Thomas P. Ketterl[†], Richard D. Gitlin[†], Peter J. Fabri[*]

[†]Department of Electrical Engineering
[*]Department of Industrial Engineering and College of Medicine
University of South Florida
Tampa, Florida 33620, U.S.A.
Email: {garrobo, calvin4, yangl}@mail.usf.edu, {ketterl, richgitlin}@usf.edu, pfabri@health.usf.edu

*Abstract*—This paper presents the proof-of-concept investigation for a miniaturized vectorcardiogram [VCG] system for ambulatory on-body applications that continuously monitors the electrical activity of the heart in three dimensions. We investigate the minimum distance between a pair of leads in the X, Y and Z axes such that the signals are distinguishable from the noise. The target dimensions for our VCG are 3x3x2 cm and, based on our preliminary results, it is possible to achieve these dimensions. The next step in our research is to build the miniaturized VCG system that includes processing, learning and communication capabilities.

## I. Introduction

Cardiac Rhythm Management (CRM) is the field of cardiovascular disease therapy that relates to the detection and treatment of abnormally fast and abnormally slow heart rhythms [1]. Current rhythm management systems manage heart activity by monitoring electric impulses generated by the heart, or via other means such as blood acceleration as it leaves the heart, and take corrective action upon sensing dangerous or potentially dangerous changes in normal heart activity. Cardiac rhythm management devices have come a long way from when they were first invented and used in the 1950s. Most such systems now have external controllability and monitoring so that the physician can check the condition of the heart through an external device, on a routine basis, typically on a monthly basis. PDAs and cell phones are used to facilitate this function. Implantable defibrillator devices (ICD), also known as Automatic Implantable Cardioverter Defibrillators (AICD), have a built-in Electrocardiogram [EGM] monitor, called an intracardiac electrogram (EGM) monitor [2].

The electrocardiogram (ECG/EKG) is an electrical representation of the polarization cycle of the heart as read by electrodes on the skin surface placed at strategic points on the body. The heart is stimulated by the natural pacemaker of the heart, which applies an electric current to the heart such that the muscles of the heart (coronary muscles) contract and expand. This helps to present an accurate picture of the health of the heart. The 12-lead ECG, referred to as the gold standard, has 4 probes that are placed on the body extremities and 6 precordial leads on the chest (near the heart).

Vectorcardiography [VCG] presents a compact three-dimensional (3D) view of the depolarization (depolarization cycle) of the heart by calculating the magnitude and direction of the electrical signals emanated from the heart [1], [3]. These vectors are used to make 3 projections of the polarization event of the heart, namely coronal (frontal), transverse (horizontal) and sagittal (vertical) plane. Vectorcardiography is presently used mostly for didactic purposes to teach students of biomedical sciences, physiological aspects of electrocardiography. In these cases, the 12-lead ECG is converted into the X, Y, and Z vectors via a 12x3 matrix transformation. The VCG and the 12-lead ECG represent the same information, albeit in different formats.

The principal advantage of the VCG is that it provides the same information as the 12-lead ECG but with fewer leads. This is achieved, as mentioned, by manipulating these orthogonal vector signals to yield a conventional ECG signal.

The VCG was created in the 1930s [4] and was viewed primarily as a teaching tool. In this paper, the VCG concept is extended to enable real-time monitoring of the hearts electrical activity of patient with a small form factor device that can be worn on the body of the patient. In section II, we present a literature review of the vectorcardiogram. In section III we describe our novel VCG as a part of a CRM system. Section IV presents the test setup and preliminary results of our approach. Finally, we present our conclusions and future research directions in section V.

## II. Literature Review

Vectorcardiogram (VCG) is a method of using a series of vectors to record the magnitude and direction of the electrical forces in electrical cardiac activities. Fig. 1 shows a view of the VCG signal from three Cartesian planes (X-Y, Y-Z and X-Z). VCG and electrocardiogram (ECG) are both reflections of the electric cardiac activities and they differ in the recording methods. It is widely accepted that VCG is an important complement and improvement to ECG and a more accurate clinical diagnosis may be obtained when combining them together. The quality and the quantity of information in the VCG is at least as comprehensive as the 12-lead ECG and may contain information that is useful in certain circumstances. VCG was not commonly applied in clinical diagnosis before because of the higher cost of the equipment. However, along with the improvement of VCG instruments and simplification of the recording method, we expect that VCG-based systems will be more widely used in clinical research and applications.

The initial concept of VCG was introduced in 1920 [5]. The author manually graphed a series of vectors to represent the electric forces of depolarization and repolarization according to the recordings of the electric cardiac activity. The first technique for recording the VCG was reported in 1936 and 1937 in [6]–[8]. The Frank system [9] was an improved system for spatial vectorcardiography which made the VCG clinically practical. It was introduced in 1956 and is still widely used at present. The method to record VCG by the Frank system is shown in Fig. 2.

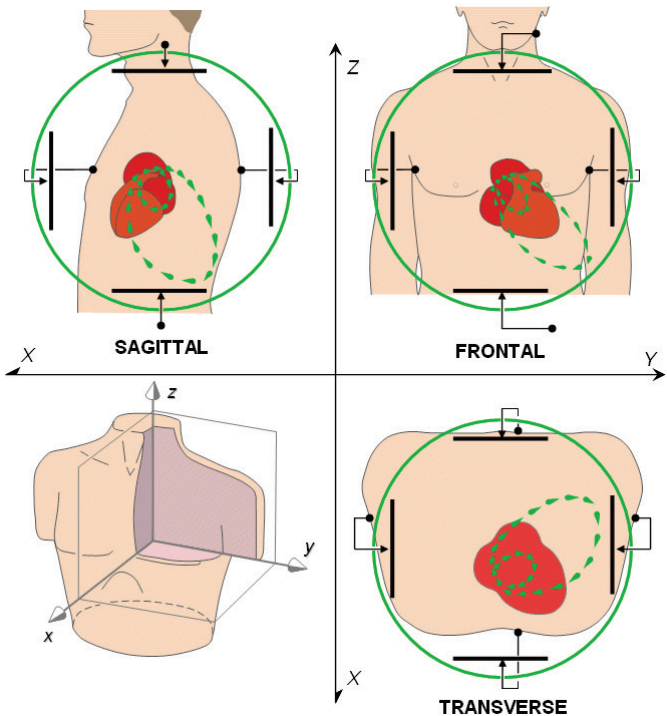

Fig. 1. The view of VCG in three Cartesian planes [3]

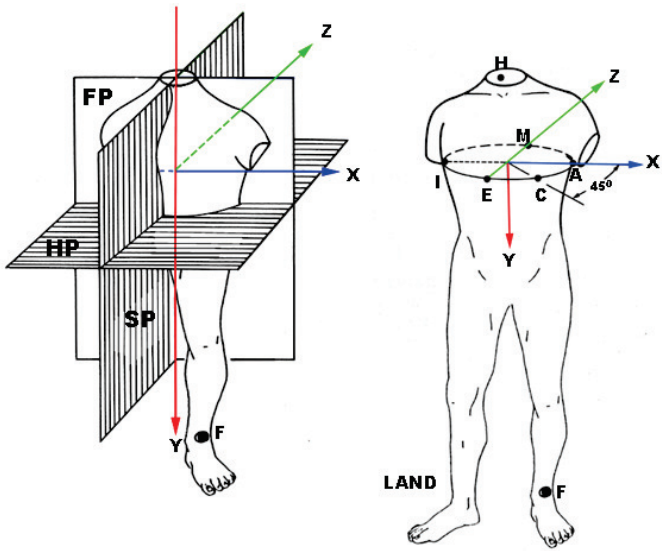

Fig. 2. Location of electrodes to perform VCG by the Frank method [9]

Some recent research shows the benefits of using the VCG in different aspects in clinical medicine field. In [10], a new and simple method is presented, which is based on the baseline VCG analysis to predict responses in cardiac resynchronization therapy (CRT) and in [11] the authors reported that the frontal plane heart depolarization vector cardiogram (HDVC) can be diagnostically applied to identify certain cardiac disorders. The authors in [12] use a method that combines several VCG variables to obtain better diagnostic accuracy of left ventricular hypertrophy (LVH), compared to the conventional ECG criteria. Furthermore, there is also considerable recent research focused on the representations or computations of VCG signals. For example, a Bayesian method which can determine a VCG, tailored to each specific patient, is proposed in [13]. In [14], the authors derive the 12-lead ECG and 3-lead VCG from 3 measured leads with 5 electrodes, which is an efficient way to obtain the ECG and VCG and has significance in acute patient care. Some research is presented in [15] regarding the statistical affine transformations between 3-lead VCG and 12-lead ECG, which is proved to be more accurate than the traditional methods such as Dowers transformation. In [16], different methods are compared in computing spatial QRS-T angles (SA) in VCG signals and it is discovered that the method which uses the Kors matrix shows superior correspondence than the others. The authors in [17] propose a modeling approach based on multi-scale adaptive basis function to characterize both temporal and spatial behaviors of VCG signals. Their proposed model shows great potential to model space-time cardiac pathological behaviors and also has some benefits in practical applications. In [18], the authors present an investigation on spatiotemporal VCG signal representation, which can be used to get a better medical interpretation and clinical applications of VCG. Their research presents both spatial and temporal characteristics of VCG signals in dynamic representation, which benefits the assessment of cardiovascular diseases.

III. VECTORCARDIOGRAM SYSTEM

A. Vector Cardiac Rhythm Monitoring System

In this paper we present a novel on-body vectorcardiogram, with a small form factor, that is part of a vector cardio rhythm monitoring system (*v*CRM). The *v*CRM has a very small form factor, such that it can be an ambulatory device and can continuously monitor the electrical activity of the heart in 3D. The main challenge is to build a small device (small separation between leads) to capture the activity of the heart in X, Y and Z dimensions. That is, the VCG uses three orthogonal systems of leads to obtain the 3D electrical representation of the heart. This miniaturized VCG can be given learning and wireless communication capabilities. The *v*CRM system is comprised of a miniaturized wireless VCG, Pacemaker (CRT/ICD), and an associated server, as shown in Fig. 3.

The *v*CRM enables comprehensive long term 24x7 information collection "BIG DATA", from an outpatient that is similar to that available from the office-based 12-lead ECG, to be continuously received and processed. This capability has never been available before.

B. Vectorcardiogram System

The first step to realize our *v*CRM system is to design and build a miniaturized VCG that can be worn on the body. To achieve this, we placed two pairs of electrodes on the body (one pair in the X-axis and another pair in the Y-axis at different distances) as illustrated in Fig. 4. For the Z-axis, we place one electrode on the chest and the other electrode on the back. As the proximity between the leads in the X and Y axes is decreased, the amplitude and wave shape of the signals will, at some point, suffer loss of amplitude and become distorted (orthogonality) and be degraded relative to that of a 12-lead ECG.

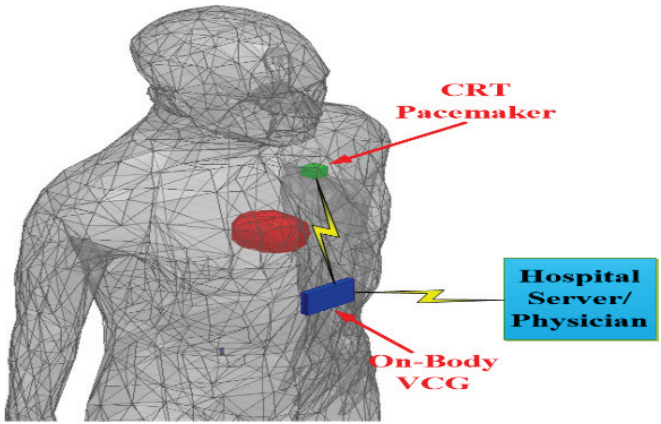

Fig. 3. Vector Cardio Rhythm Monitoring System (*v*CRM)

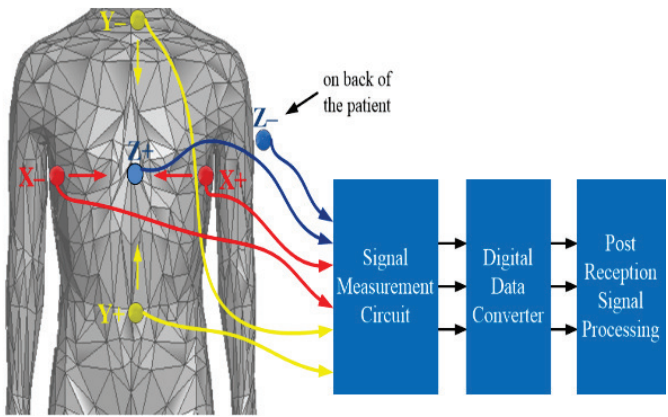

Fig. 4. Location of the electrodes on the body in the X and Y axes at different distances and breadboard to record the signals

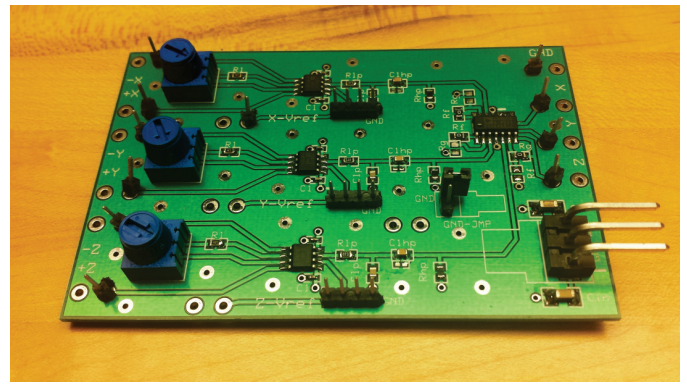

Fig. 5. *v*CRM Breadboard Recording System

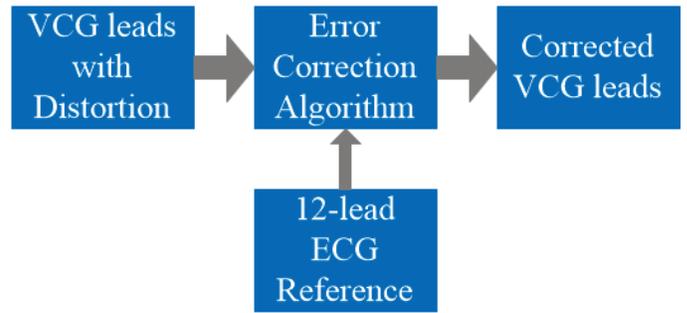

Fig. 6. VCG calibration

We built the breadboard version of the VCG to make measurements and experiment with signal processing, as shown in Fig. 5, before we construct the miniaturized VCG. The goal of the miniaturized VCG design is to fit all the electronics into a 3x3x2 cm package. The breadboard (Signal Measurement Circuit in Fig. 4) is a 4-layer PCB used to connect the electrodes and collect the electrical signals coming from the heart. The dimensions are 9.65x6.35 cm. The breadboard contains one differential amplifier for each dimension (X, Y and Z). Those signals are passed to the computer through a data acquisition card (Digital Data Converter in Fig. 4) for post processing.

From preliminary results, we saw that 60 Hz noise is introduced in the measurements due to the electrical power system. Because of this, the Post Reception Signal Processing in Fig. 4 is used to eliminate this unwanted 60 Hz noise signal. Additionally, the signal processing is used to re-orthogonalize the measured data, and resolve any other issues that have arisen. That is, at the time of fitting the VCG device to the patient, a 12-lead ECG can be fitted for calibration, as illustrated in Fig. 6. Both the VCG signals and the 12-lead ECG signals are fed into an algorithm that determines the underlying transfer function and the appropriate signal processing. This transfer function will differ from person to person and must be customized for each patient. As a result, the VCG will be part of the Personalized Medicine revolution in medicine. The following section presents preliminary results of our novel vectorcardiogram system.

## IV. PRELIMINARY RESULTS

We ran several preliminary tests to measure the electrical activity of the heart for different distances, and obtained data for different distances between two leads in the X and Y axes, as shown in Fig. 7 and 8, respectively. This information is useful to determine the minimum distance between two leads such that a satisfactory diagnosis can be done. As was mentioned in the previous section, we placed an electrode on the chest (Z) and other one in the back (Z+) for the Z axis.

As we can see in Figs. 7 and 8, the closer the distance between the leads, the smaller the amplitude of the signal and the larger the noise. Fig. 9 presents the measured and filtered signal for the X-axis at a distance between the pair of electrodes equal to 10.67 cm. As we can see, the post - processing block helps to remove the 60 Hz noise and other unwanted signals.

The filtered signals for the X, Y and Z axes are shown in Fig. 10. The amplitude of the signal in the Z-axis is larger than the X and Y-axes because the distance is very large compared to the distance in the X and Y-axis. These preliminary results, demonstrate that it is probably quite possible to miniaturize the VCG for on-body ambulatory applications and achieve our targeted dimensions (3x3x2 cm).

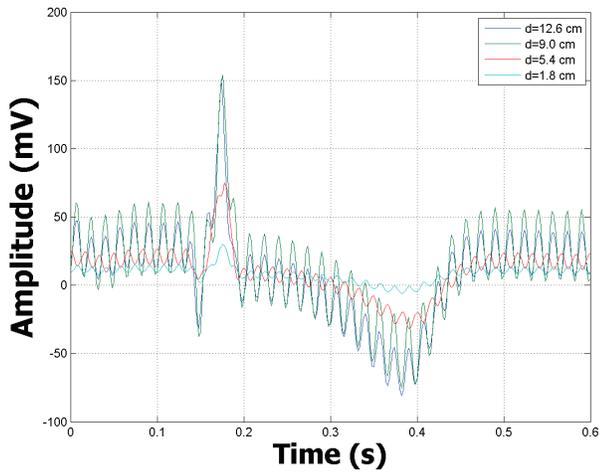

Fig. 7. Measured signals for different distances in the X axis

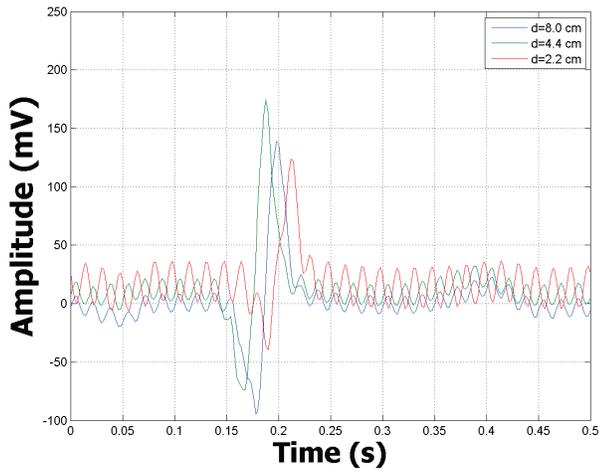

Fig. 8. Measured signals for different distances in the Y axis

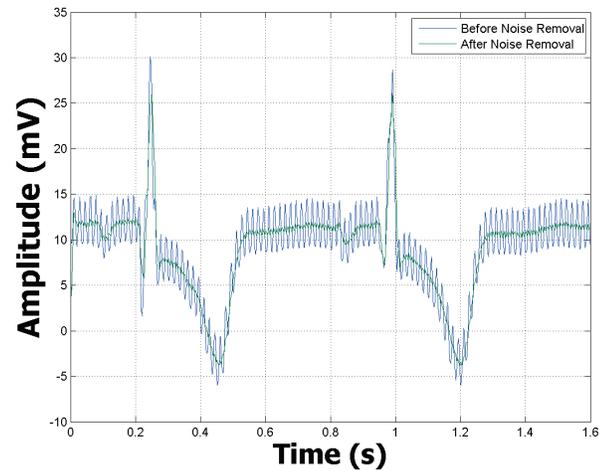

Fig. 9. Measured and filtered signal in the X axis

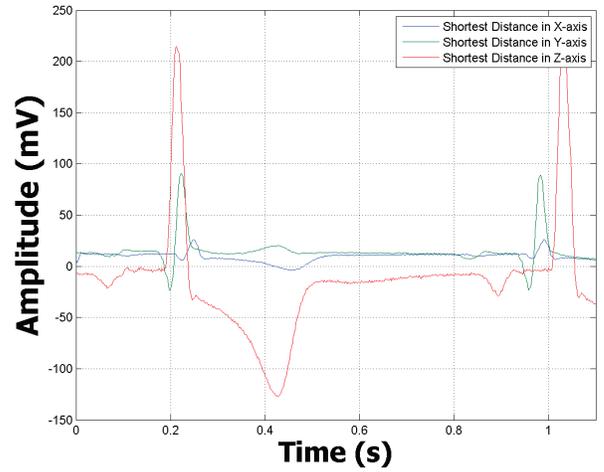

Fig. 10. Filtered signals for the smaller distances in the X and Y axis

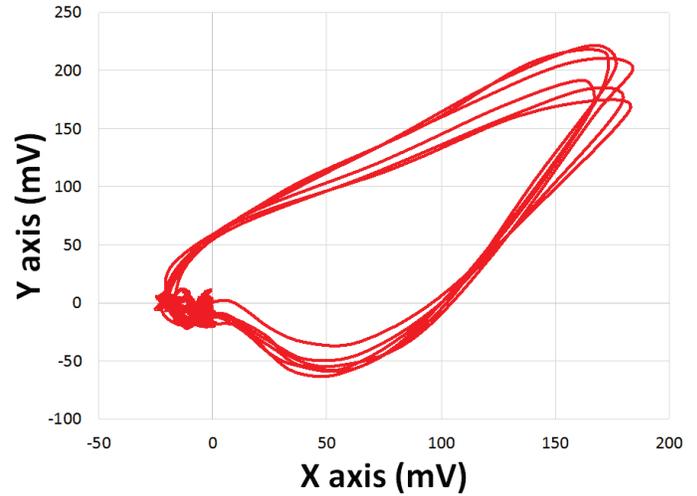

Fig. 11. VCG Loops Plotted in X-Y plane

Figures 11, 12, and 13 are a graphical illustration showing the VCG loops plotted in X-Y, Y-Z and X-Z planes, respectively. Figure 14 shows the VCG loops in 3D.

## V. CONCLUSIONS AND FUTURE RESEARCH

The Vector Cardio Rhythm Monitoring (*v*CRM) system will be a breakthrough and disruptive technology allowing long term and continuous remote monitoring of patients electrical heart activity. The implications are potentially profound and include 1) providing a less expensive device compared to the 12-lead ECG system [the gold standard]; 2) continuous remote tele-monitoring of a patient becomes possible; 3) the current Holter monitor system can be replaced; 4) ICU cardiac monitoring becomes much less complex and costly; 5) physicians can potentially send people/patients home earlier because the doctors can have continuous remote monitoring of the patients.

The main component of our *v*CRM system is the miniaturized VCG and in this paper, we have presented proof of concept experimental results that strongly suggest that our novel

VCG can be realized. Moreover, in this paper we concentrated solely on collecting and processing the signals of the electrical activity of the heart. The learning system and communication capabilities are being added to our miniaturized VCG system such that it can learn the patterns of the collected data and make decisions that are sent to the pacemaker (ICD/CRT).

The next steps in our research are to build the miniaturized VCG that includes processing, learning and wireless communication capabilities.

ACKNOWLEDGMENT

This research was supported in part by Jabil Inc., the Florida 21st Century Scholars program and the Florida High Tech Corridor Matching Grants Research Program.

REFERENCES

[1] *Comprehensive electrocardiology: theory and practice in health and disease*, 1st ed.   Pergamon Press.
[2] A. Kachenoura, F. Poree, A. Hernandez, and G. Carrault, "Surface ECG reconstruction from intracardiac EGM: a PCA-vectorcardiogarm method."   IEEE, pp. 761–764.
[3] J. Malmivuo and R. Plonsey, *Bioelectromagnetism: Principles and Applications of Bioelectric and Biomagnetic Fields*.   Oxford University Press.
[4] G. E. Burch, "The history of vectorcardiography," no. 5, pp. 103–131.
[5] MANN H, "A method of analyzing the electrocardiogram," vol. 25, no. 3, pp. 283–294.
[6] F. Schellong, "Elektrographische diagnostik der herzmuskelerkrankungen," no. 48, pp. 288–310.
[7] F. Wilson and F. Johnston, "The use of the cathode-ray oscillograph in the study of the monocardiogram."
[8] W. Hollmann and H.E. Hollmann, "Neue elektrokardiographische untersuchungsmethoden," no. 29, p. 546.
[9] E. Frank, "An accurate, clinically practical system for spatial vectorcardiography," vol. 13, no. 5, pp. 737–749.
[10] T. Schau, W. Koglek, J. Brandl, M. Seifert, J. Meyhfer, M. Neuss, G. Grimm, R. Bitschnau, and C. Butter, "Baseline vectorcardiography as a predictor of invasively determined acute hemodynamic response to cardiac resynchronization therapy," vol. 102, no. 2, pp. 129–138.
[11] D. N. Ghista, U. R. Acharya, and T. Nagenthiran, "Frontal plane vectorcardiograms: Theory and graphics visualization of cardiac health status," vol. 34, no. 4, pp. 445–458.
[12] Z. Loring, C. W. Olson, C. Maynard, and N. Hakacova, "Modeling vectorcardiograms based on left ventricle papillary muscle position," vol. 44, no. 5, pp. 584–589.
[13] R. Vullings, C. Peters, I. Mossavat, S. Oei, and J. Bergmans, "Bayesian approach to patient-tailored vectorcardiography," vol. 57, no. 3, pp. 586–595.
[14] D. M. Schreck and R. D. Fishberg, "Derivation of the 12-lead electrocardiogram and 3-lead vectorcardiogram," vol. 31, no. 8, pp. 1183–1190.
[15] D. Dawson, H. Yang, M. Malshe, S. T. Bukkapatnam, B. Benjamin, and R. Komanduri, "Linear affine transformations between 3-lead (frank XYZ leads) vectorcardiogram and 12-lead electrocardiogram signals," vol. 42, no. 6, pp. 622–630.
[16] C. A. Schreurs, A. M. Algra, S.-C. Man, S. C. Cannegieter, E. E. van der Wall, M. J. Schalij, J. A. Kors, and C. A. Swenne, "The spatial QRS-t angle in the frank vectorcardiogram: accuracy of estimates derived from the 12-lead electrocardiogram," vol. 43, no. 4, pp. 294–301.
[17] G. Liu and H. Yang, "Multiscale adaptive basis function modeling of spatiotemporal vectorcardiogram signals," vol. 17, no. 2, pp. 484–492.
[18] H. Yang, S. T. Bukkapatnam, and R. Komanduri, "Spatiotemporal representation of cardiac vectorcardiogram (VCG) signals," vol. 11, no. 1, p. 16.

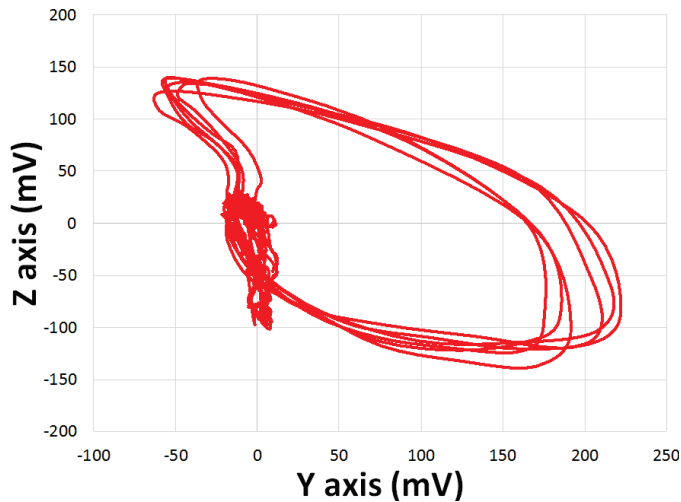

Fig. 12.   VCG Loops Plotted in Y-Z plane

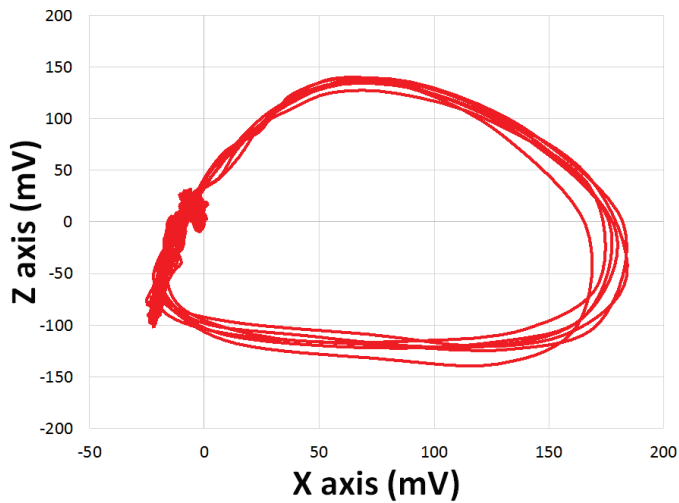

Fig. 13.   VCG Loops Plotted in X-Z plane

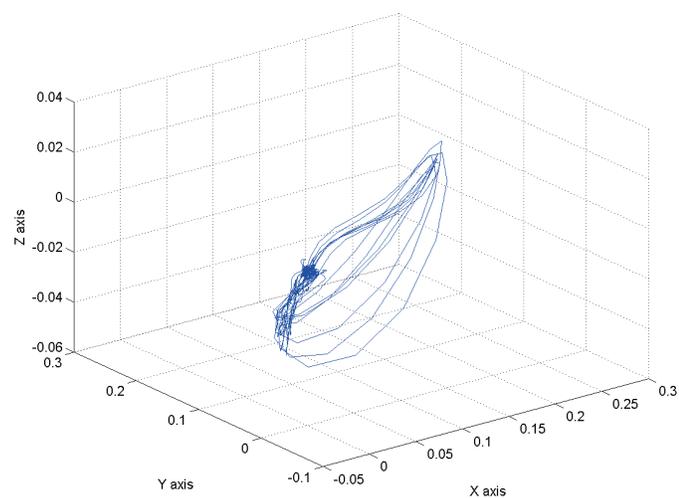

Fig. 14.   VCG Loops Plotted in Three Dimensions